\documentclass[lettersize,journal]{IEEEtran}
\usepackage[utf8]{inputenc}
\IEEEoverridecommandlockouts
\usepackage{cite}
\usepackage{amsmath,amssymb,amsfonts}
\usepackage{algorithmic}
\usepackage{graphicx}
\usepackage{textcomp}
\usepackage{xcolor}
\usepackage{algorithm}
\usepackage[inline]{enumitem}
\usepackage{pdflscape}
\usepackage{esint}
\usepackage{subfig}
\begin{document}

\title{SLA Decomposition for Network Slicing:\\ A Deep Neural Network Approach
}

% \author{\IEEEauthorblockN{Cyril Shih-Huan Hsu}
% \IEEEauthorblockA{\textit{Informatics Institute} \\
% \textit{University of Amsterdam}\\
% Amsterdam, The Netherlands \\
% s.h.hsu@uva.nl}
% \and
% \IEEEauthorblockN{Danny De Vleeschauwer}
% \IEEEauthorblockA{\textit{Nokia Bell Labs} \\
% % \textit{Nokia Bell Labs}\\
% Antwerp, Belgium \\
% danny.de\_vleeschauwer@nokia-bell-labs.com}
% \and
% \IEEEauthorblockN{Chrysa Papagianni}
% \IEEEauthorblockA{\textit{Informatics Institute} \\
% \textit{University of Amsterdam}\\
% Amsterdam, The Netherlands \\
% c.papagianni@uva.nl}
% }

\author{
\IEEEauthorblockN{Cyril Shih-Huan Hsu,
Danny De Vleeschauwer\IEEEauthorrefmark{1} and
Chrysa Papagianni
}
\thanks{Cyril Shih-Huan Hsu and Chrysa Papagianni are with Informatics Institute, University of Amsterdam, The Netherlands (email: s.h.hsu@uva.nl, c.papagianni@uva.nl).}% <-this % stops a space
\thanks{\IEEEauthorrefmark{1}Danny De Vleeschauwer is with Nokia Bell Labs, Antwerp, Belgium (e-mail: danny.de\_vleeschauwer@nokia-bell-labs.com.}
\thanks{
Work partially funded by EC-funded projects  H2020 DAEMON (grant no. 101017109) and HORIZON SNS JU DESIRE6G (grant no. 101095890).}
}

\maketitle
%{\color{red}
\begin{abstract}
For a network slice that spans multiple technology and/or administrative domains, these domains must ensure that the slice's End-to-End (E2E) Service Level Agreement (SLA) is met. 
%5G is a multi-service network supporting a range of verticals with a diverse set of requirements. Network slicing enables the creation and operation of multiple logical networks over the shared network infrastructure, tailored to the requirements of a particular service type with agreed upon Service Level Agreement (SLA). A network slice may span different parts of the network (i.e., access, core, and transport network) and could be deployed across multiple operators and infrastructure providers. 
Thus, the E2E SLA should be decomposed to partial SLAs, assigned to each of these domains. Assuming a two-level management architecture consisting of an E2E service orchestrator and local domain controllers, we consider that the former is only aware of historical data of the local controllers' responses to previous slice requests, and captures this knowledge in a risk model per domain.
%assume a two-level management architecture consisting of an end-to-end service orchestrator responsible for the lifecycle management of the network service and domain controllers that are in charge of instantiating parts of the slice in their respective domains. In this context the orchestrator is responsible for decomposing the SLA, without detailed knowledge of the state of the resources in each of the domains. The orchestrator is only aware of each domain controller's response to previous requests and captures this knowledge in a risk model associated with the domain. 
In this study, we propose the use of Neural Network (NN) based risk models, using such historical data, to decompose the E2E SLA.
Specifically, we introduce models that incorporate monotonicity, applicable even in cases involving small datasets. An empirical study on a synthetic multi-domain dataset demonstrates the efficiency of our approach.
%an approach for decomposing the E2E SLA adopting neural network-based risk models. We discuss approaches that use a monotonicity prior, such that the SLA can be decomposed even when the number of historical data is low. An empirical study on a synthetic multi-domain dataset demonstrates the efficiency of our approach.
\end{abstract}
%}

\section{Introduction}
Continuing the trend from 5G systems, the future mobile network is expected to be a multi-service network supporting multiple vertical industries with a diverse set of requirements. 
%The current generation of mobile communications (5G) specifies a multi-service network supporting various vertical industries with a diverse set of service requirements. This e
Network slicing, introduced in 5G, enables the creation and operation of multiple logical networks over the shared infrastructure, tailored to the requirements of services with agreed upon SLAs. SLAs are provider-customer contracts setting the expected quality, performance, and availability of a service in terms of measurable targets such as throughput, latency, reliability etc., known as Service-Level Objectives (SLOs).

A network slice may span different parts of the network (i.e., access, transport and core network) and could be deployed across multiple infrastructure providers. The deployment of the E2E service across a set of domains must meet the agreed upon SLOs. To this end, the E2E SLA associated with the slice should be decomposed into partial SLOs assigned and supported by each of these domains. % that can be effectively used to perform resource allocation.
Decomposing the E2E SLA in the underlying domains’ requirements is an inevitable step in resource allocation \cite{su2019resource}. AI-assisted SLA decomposition is considered key to automating 6G complex business processes \cite{10011552}.
%Thus, SLA decomposition has been acknowledged as a significant challenge in resource allocation for network slicing \cite{Vleeschauwer21_SLAdecomposition}.

\begin{figure}[h]
 \centering     \captionsetup{justification=justified,singlelinecheck=off}
\includegraphics[width=0.8\columnwidth]{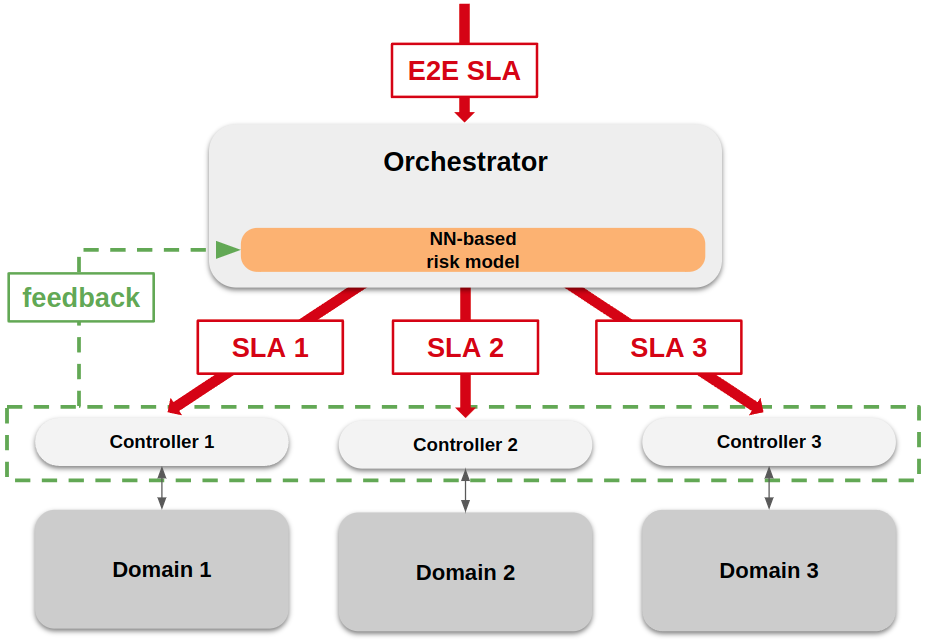}
% \justifying
\caption{Network slicing management and orchestration system}
\label{fig:NESMOS}
\vspace{-0.5em}
\end{figure}

Similar to  \cite{Vleeschauwer21_SLAdecomposition}, we assume a two-level management architecture consisting of an E2E service orchestrator responsible for network service lifecycle management, and local domain controllers that are in charge of instantiating parts of the network slice in their respective domains (Fig.~\ref{fig:NESMOS}). % depicts the management framework which comprises an end-to-end service orchestrator, domains, and their controllers. The problem we tackle in this work lies in the interaction between orchestrator and domain controllers. 
The  orchestrator determines the SLA decomposition for the incoming service request, while the domain controllers perform admission control within their  domains, prior to resource allocation required for E2E slice deployment. %When a portion of the end-to-end SLA is offered by the orchestrator, the controllers have to decide to either accept or reject the request. 
We  assume that the orchestrator has no knowledge of the state of the infrastructure %for the underlying domains,
at the moment that SLA decomposition is performed. However, we consider that admission control information (i.e., request acceptance or rejection) from each domain is accessible to the orchestrator. Thus, it can make informed decisions using domain-specific risk models employing such data.

Existing approaches for SLA decomposition mainly employ heuristics, considering state information periodically reported by the underlying domains\cite{su2019resource}.
Authors in~\cite{9165317} propose an E2E SLA decomposition system, using supervised machine learning algorithms. %However, the models are designed for a fixed number of domains.
In our previous work~\cite{Vleeschauwer21_SLAdecomposition}, the problem is addressed in three steps: 
($i$) we formulate the decomposition problem as a function of the SLA acceptance probability per domain, under the constraints set by the E2E SLOs;
($ii$) we use a parameter-free risk model per domain to estimate these probabilities, constructed by observing the response of domain controllers to previous requests. We formulate the risk model estimation as a likelihood maximization problem, under monotonicity constraints stemming from the nature of the decomposition problem and employ Sequential Quadratic Programming (SQP) to solve it.
($iii$) Using the risk models, we apply an initial exhaustive search followed by SQP to solve the decomposition problem for each request.
However, the estimation of a parameter-free risk model is non-linear, hence computationally intensive. 

In this paper we introduce risk models based on deep NNs to support the SLA decomposition process.
%($i$) maximizing the likelihood of the samples in the training dataset under the monotonicity constraint, which determines the acceptance probability of each SLA; ($ii$) predicting the acceptance probability with the parameter-free model: for an incoming SLA, finding its upper and lower bounds from the samples in the training dataset, averaging the two associated acceptance probabilities, and reporting the result as the predicted acceptance probability; ($iii$) employing Sequential Quadratic Programming to determine the optimal decomposition, given the risk model acquired in the previous step. In this work, we propose alternative methods for step ($i$) and ($ii$).
%The main contribution of this paper is an approach for SLA decomposition introducing risk models based on Deep Neural Networks (NNs). 
In general, we want to make as few assumptions as possible to obtain the risk model per domain~\cite{Vleeschauwer21_SLAdecomposition}. %In fact, the only constraint we impose is that the monotonic property is obeyed. 
To this end, we propose methods that allow NN-based risk models to approximate monotonic functions without compromising their expressiveness. Most of these models perform robustly even with small datasets. Furthermore, they are capable of learning from a more complex dataset with a constant inference time, %(i.e., that does not increase with the number of samples), 
thus SLA decomposition improves accordingly, increasing the accuracy and scalability of the approach. Regarding the time required for model construction, it scales linearly with the number of samples, while  the approach in \cite{Vleeschauwer21_SLAdecomposition} scales polynomially. This make them also appropriate for online-learning risk models, enabling adaptation in a more dynamic environment.

%The rest of this paper is organized as follows. 
Section \ref{sec:problem} provides an overview of the SLA decomposition problem. In Section \ref{sec:methods} we present the proposed deep NN-based risk models. In Section \ref{sec:performance} we describe the experimental setup while in Section \ref{sec:results} we evaluate the problem using the proposed models. Section \ref{sec:conclusion} presents our conclusions.

\section{Problem Description}
\label{sec:problem}

\subsection{SLA Decomposition}
The E2E SLA $s_{e2e}$ is described as a vector of SLOs with respect to certain performance metrics~\cite{Vleeschauwer21_SLAdecomposition}. The values of the SLO vector represent the constraints defined in SLA in order; for example, an SLA comprises an E2E delay and throughput, i.e., $s_{e2e}=(\tau_{e2e}, \theta_{e2e})$. This vector indicates that the corresponding network slice has to be operated in the way that its E2E delay $\tau$ and throughput $\theta$ fulfill the constraints given by the SLO vector, i.e., $\tau \leq \tau_{e2e}$, and $\theta \ge \theta_{e2e}$.
Assuming that the network slice spans $N$ domains, where $n=\{1, 2,..., N\}$, we define the vector $s_n$ that represents the $n$-th domain SLOs, and the equation $s_{e2e}=G(s_1, s_2,..., s_N)$ that describes how per domain SLOs form the E2E objective $s_{e2e}$. For instance, the E2E delay is the sum of all delays for the involved domains, and the E2E throughput is the minimum throughput from all domains, i.e., $\tau_{e2e}=\Sigma_{n=1}^{N}\tau_n$ and $\theta_{e2e}=\min\{\theta_1, \theta_2,..., \theta_N\}$.

The ability of a domain to support a partial SLA $s_n$, is captured by a risk model. The risk models of all involved domains can be used in the SLA decomposition process.
%Now, if we have the risk model that adequately captures the probability of how likely a domain is able to support the SLA portion $s_n$, the decomposition can be made accordingly. 
The risk model is defined as $-\log P_n(s_n)$, where $P_n(s_n)$ represents the probability that a request in the $n$-th domain with  $s_n$ SLOs is accepted. Assuming each domain makes decisions independently, then the E2E acceptance probability is the product of the acceptance probabilities of all individual domains. Therefore, the E2E decomposition can be formulated as an optimization problem that minimizes the objective~(\ref{eq:obj}) under the constraints~(\ref{eq:composition}) as follows:

\begin{equation} \label{eq:obj}
\min\limits_{\bf{s}} -\sum_{n}^{N} \log P_n(s_n)
\end{equation}
\begin{equation} \label{eq:composition}
\mbox{s.t.}\ s_{e2e}=G(s_1, s_2,..., s_N)
\end{equation}

\noindent Furthermore, if a parameterized neural network $\mathcal{F}_n$ is utilized as the risk model, where $\mathcal{F}_n(s_n)=P_n(s_n)$, the objective function in Eq.~\ref{eq:obj} can be rewritten as:
\begin{equation}
-\sum_{n}^{N} \log \mathcal{F}_n(s_n)
\end{equation}

\noindent  The goal of the decomposition policy is to minimize the overall risk of rejecting the corresponding request with SLOs $s_n$ for each participant $n$-th domain.

\vspace{-1em}
\subsection{Determining Neural Network-based Risk Models}
 Based on our previous work~\cite{Vleeschauwer21_SLAdecomposition}, we determine the risk model per domain. 
 Given a new request with specific SLOs $s$ (we omit the domain subscript $n$ as we focus on a single domain), a controller has to decide whether or not to accept the request. To make such decision,  the controller  has to consider also the state $\omega$ of the infrastructure at the moment. This state is defined by the utilization and loads on the links and servers, the delays incurred over network links and the calculated backup paths, etc. 
It is also determined by the randomness of the SLAs of all previous incoming  requests as well as the decisions taken by the domain. Therefore, certain $(s, \omega)$ pairs will lead to acceptance and others to a rejection.  Note that the domain controller has detailed information about the state and the impact of the decision it will make, but it is not the case for the orchestrator.  Despite the fact that the underlying decision-making process of the controller is deterministic, the orchestrator still experiences this as stochastic, because it is unaware of the state $\omega$ of the infrastructure.

However, the acceptance probability $P$ can be modeled by observing the outcomes of the domain controller to the previous requests. Given $K$ observations $\{(x_1, y_1), (x_2, y_2),..., (x_K, y_K)\}$, where $k=\{1, 2,..., K\}$, $x_k$ is the proposed SLO to the domain, and $y_k\in\{0, 1\}$ represents its associated response, i.e., being rejected or accepted, the acceptance probabilities for these SLA vectors $\{x_1, x_2,..., x_K\}$ can be modelled by parameterised neural networks $\mathcal{F}$ via maximizing the following likelihood:
\begin{equation} \label{eq:risk}
\sum_{i=1}^{K}[y_i*\log(\mathcal{F}(x_i)) + (1-y_i)*\log(1-\mathcal{F}(x_i))],
\end{equation}
where $\mathcal{F}(x_i) \in [0, 1]$ is the estimated acceptance probability of the SLO $x_i$ predicted by the neural network.
Note that $x$ is a realization of the variable $s$ i.e., the SLO vector per domain.

\vspace{-1em}
\subsection{Monotonicity}
The acceptance probability based on the ground truth follows a partial order relation \cite{Vleeschauwer21_SLAdecomposition}, which incorporates the notion of
a stricter SLA, i.e., given a set of $K$ SLOs $S=\{x_1, x_2,..., x_K\}$, the acceptance probability has the following property: %(the domain subscript $n$ is omitted as we now focus on a single domain):
\begin{equation} \label{eq:partial}
\forall x_i, x_j \in S, P(x_i) \leq P(x_j) \quad \textrm{if} \quad x_i \preceq x_j,
\end{equation}
The property indicates that a stricter SLO $x_i$ is less likely to be accepted. Note that this is a partial instead of total order, because only a part of the vectors stand in such a relation to each other. As an example, an SLA is characterised by a (delay, throughput) tuple, the vector $x_i=(\tau_i, \theta_i)$ is stricter than the other vector $x_j=(\tau_j, \theta_j)$, i.e., $x_i \preceq x_j$ if and only if $\tau_i \leq \tau_j$ and $\theta_i \ge \theta_j$. Hence, in this case Eq.~\ref{eq:partial} states that an SLA with lower delay and higher throughput requirements is less likely to be accepted.

In order to benefit from this prior knowledge,  the NN which we utilize as the risk model should also be monotonic.  Many research efforts have been carried out to incorporate monotonicity into NNs while maintaining their expressiveness~\cite{liu2020certified, NIPS1997_83adc922}.
% The existing approaches for learning monotonic NNs can be roughly categorized into two kinds; (i) special neural architectures and (ii) heuristic regularization.
% The former methods usually guarantee monotonicity, yet tend to be difficult to implement or train. The latter ones are typically simple to apply to an arbitrary, off-the-shelf neural network, however monotonicity is not ensured. We introduce several approaches that belong to the two categories in Section~\ref{sec:methods} to learn each involved domain's risk model for SLA decomposition.
We introduce six approaches in Section~\ref{sec:methods} to learn each involved domain's risk model for SLA decomposition.
\section{Approaches}
\label{sec:methods}

In this section, the six NN-based methods are proposed, to construct a monotonic risk model for each domain. Table~\ref{table:comp} summarises the characteristics of all proposed methods.

\noindent \textbf{Vanilla Neural Network.}
A plain NN is used as the baseline model. The goal is to find a parameterized neural network $\mathcal{F}$ that minimizes the binary cross entropy loss:
\begin{equation} \label{eq:bce}
L_{BCE} = -\frac{1}{K}\sum_{i=1}^{K}[y_i*\log(\widehat{y_i}) + (1-y_i)*\log(1-\widehat{y_i})]
\end{equation}
where $\widehat{y_i}=\mathcal{F}(x_i)$ is the prediction by the neural network.

\noindent \textbf{Regularised Neural Network.}
One of the sufficient conditions that guarantees the monotonicity of a NN is to have all its weights non-negative~\cite{NIPS1997_83adc922, https://doi.org/10.1111/j.1540-5915.1993.tb00462.x}. To achieve this, on top of the settings of the vanilla NN, we penalize the negative weights, as an additional regularisation loss during training. Given a set of $J$ weights $W=\{w_1, w_2,.., w_{J}\}$, this loss is defined as:
\begin{equation}
L_{Reg} = \sum_{j=1}^{J}\delta(w_j)*w_j^2,\;
    \delta(w)= 
\begin{cases}
    1,& \text{if } w< 0\\
    0,              & \text{otherwise}.
\end{cases}
\end{equation}
% \[
%     \delta(w)= 
% \begin{cases}
%     1,& \text{if } w< 0\\
%     0,              & \text{otherwise}.
% \end{cases}
% \]
%\begin{equation}

%\end{equation}
\noindent The regularisation loss is considered together with the binary cross entropy loss in Eq.~\ref{eq:bce} during the optimization process. The total loss is then defined as:
\begin{equation} \label{eq:reg}
L_{Total} = L_{BCE} + k*L_{Reg}
\end{equation}
with the parameter $k$ to balance between two loss terms.  Note that the monotonicity is not strictly obeyed with this approach, as the regularisation loss is a soft constraint.

\noindent \textbf{Absolute Weight Transformation (AWET).}
Following the incentive of a NN with non-negative weights, we consider a NN with the absolute value transformation applied to its weights before the forward computation. Thus instead of calculating $\textbf{w}*\textbf{x}+b$, we use $\mbox{abs}(\textbf{w})*\textbf{x}+b$, where $\textbf{w}$ is the weight vector, $\textbf{x}$ is the input vector, $b$ is the bias and $\mbox{abs}(.)$ being the absolute value function. The weights are forced to be non-negative after the transformation, thus the optimization algorithm is allowed to update freely $w$ during training while maintaining the non-negative weights constraint. The corresponding back-propagation is handled automatically by modern deep learning libraries, thus no additional modification is required. %Although the monotonicity is ensured, previous research efforts~\cite{liu2020certified} have pointed out that enforcing architecture-wise modifications could yield a fairly restrictive subset of monotonic functions.

\noindent \textbf{Mini-batch Order Loss (MOL).}
Considering the pair-wise relations between samples within a mini-batch has been successfully used for training  in recent years~\cite{zhao2016energy}. We propose the mini-batch order loss, where if the given ordered relation is not followed for all possible pairs within a mini-batch, an extra loss term is added to penalize the violations. Given a mini-batch with $M$ samples $B=\{(x_1,y_1), (x_2,y_2),..., (x_M,y_M)\}$, the mini-batch order loss is defined as:

\begin{equation}
L_{MOL} = \sum_{i}^{M}\sum_{j\neq i}^{M}S(x_i,y_i,x_j,y_j),
\end{equation}
\[
    S(x_i,y_i,x_j,y_j)= 
\begin{cases}
    \max(y_i-y_j, 0),& \text{if } x_i \preceq x_j\\
    0,              & \text{otherwise}.
\end{cases}
\]
The purpose of the loss is to encourage the predictions for a pair of samples to follow the order relation of their inputs, if the order exists. When the predictions disobey the order, for instance, $x_i \preceq x_j$ but $y_i > y_j$, a loss occurs to correct the order relation between $y_i$ and $y_j$ until $y_i \leq y_j$.
In conjunction with the binary cross entropy loss in Eq.~\ref{eq:bce}, the total loss becomes:
\begin{equation} \label{eq:mol}
L_{Total} = L_{BCE} + k*L_{MOL}
\end{equation}
with the parameter $k$ to balance between two loss terms.% Note that the mini-batch order loss is also a soft constraint, namely the monotonicity is not rigorously assured.

\noindent \textbf{Conflicting Sample Elimination (CSE).}
Another factor that could induce non-monotonic behavior to the model is that the training data itself violates the order relation. %, due to the intrinsic stochasticity of the sampling process. 
Therefore, preprocessing of the data becomes crucial. To address this, we propose the gradual elimination of samples conflicting the most with other samples.  The algorithm is provided in Alg.~\ref{alg:cse}.

\begin{algorithm}
  \caption{Conflicting Sample Elimination} \label{alg:cse}
  \begin{algorithmic}[1]
    \STATE create table $T$ and record the number of conflicting samples per sample
    \STATE remove the sample with the largest number of conflicting samples
    \STATE update table $T$
    \STATE repeat step 2 and 3 until no conflicting samples exist for all samples
  \end{algorithmic}
\end{algorithm}

\begin{table}[t!]
\caption{Comparison of the proposed methods}
\label{table:comp}
\resizebox{\columnwidth}{!}{%
\begin{tabular}{ c c c c c }
 \hline
 \textbf{Method} & \textbf{strict monotonicity} & \textbf{weight restriction} & \textbf{data refinement} & \textbf{auxiliary loss}\\
 \hline
 Vanilla   & - & - & - & -\\
 Reg.    & - & \textbf{\checkmark} & - & \textbf{\checkmark}\\
 AWET    & \textbf{\checkmark} & \textbf{\checkmark} & - & -\\
 MOL      & - & - & - & \textbf{\checkmark}\\
 CSE        & - & - & \textbf{\checkmark} & -\\
 PO    & - & - & \textbf{\checkmark} & -\\
 \hline
\end{tabular}
}
\vspace{-2em}
\end{table}

 For example, given  a set of samples $S=\{s_1:(3, 1), s_2:(2, 5), s_3:(1, 2)\}$, we can see that $s_1$ and $s_2$ are conflicting samples, as $3\succeq2$ (here we define $\succeq$ as $>$) yet $1<5$, while $s_2$ and $s_3$ have no conflict because $2\succeq1$ and $5>2$. Following the same rule, $s_1$ and $s_3$ are mutual conflicting. Based on the current set $S$, table $T$ is created to track the number of conflicting samples. At step $1$, $T_1=\{s_1=2, s_2=1, s_3=1\}$. Next, the sample that has the most conflicting samples, $s_1$, is removed. The set $S_1$ becomes $S_2=\{s_2:(2, 5), s_3:(1, 2)\}$, and the table $T_1$ is updated to $T_2=\{s_2=0, s_3=0\}$. The algorithm terminates when no conflicting samples exist. %The model trained with this approach is likely to maintain the monotonicity, although the property is still not guaranteed.
 
%  Another example shown in Fig.~\ref{fig:egcse} demonstrates the data distribution without (Fig.~\ref{fig:egcse}a) and with (Fig.~\ref{fig:egcse}b) applying CSE, where red and green dots represent rejected and accepted requests, respectively.

% \begin{figure}[H]
% \captionsetup{justification=justified, singlelinecheck=off}
% \subfloat[Without CSE]{\includegraphics[width=4cm]{Figures/cse_before.png}}\qquad
% \subfloat[With CSE]{\includegraphics[width=4cm]{Figures/cse_after.png}}\qquad
% \caption{Example of Conflicting Sample Elimination} \label{fig:egcse}
% \end{figure}

 Given a dataset with $N$ samples and $M$-dimension features, the time complexity of step 1 in Alg.~\ref{alg:cse} is $\mathcal{O}(M*N^2)$, $\mathcal{O}(N^2)$ for the one-time sorting, $\mathcal{O}(1)$ for step 2, and $\mathcal{O}(N)$ for step 3 with a hash table. The overall time complexity of Alg.~\ref{alg:cse} is $\mathcal{O}(M*N^2+N^2+N*(1+N)) = \mathcal{O}(M*N^2)$.

\noindent \textbf{Probability Optimization (PO).}
The method proposed in~\cite{Vleeschauwer21_SLAdecomposition} determines the acceptance probabilities $P=\{p_1, p_2,..., p_K\}$ per sample by maximizing the likelihood:
\begin{equation} \label{eq:popt}
\sum_{i=1}^{K}[y_i*\log(p_i) + (1-y_i)*\log(1-p_i)]
\end{equation}
given the constraints in Eq.~\ref{eq:partial}. The  probability values are inferred from a parameter-free risk model. We leverage this technique for pre-processing, by replacing the binary labels of the dataset with the obtained probabilities $P$, i.e., the dataset $D=\{(x_1,y_1), (x_2,y_2),..., (x_K,y_K)\}$ becomes $D^{\prime}=\{(x_1,p_1), (x_2,p_2),..., (x_K,p_K)\}$, where $y_i$ is a binary value and $p_i$ is a probability. Finally, $D^{\prime}$ is used for training. % the NNs.

% {\color{red}[todo: make a comparison table that shows pros and cons (or properties) for the proposed approaches, with description.]}

%Additionally, we propose the redundant dependency pruning to speedup the optimization for maximizing the objective in~\ref{eq:popt} under constraints in~\ref{eq:partial}. Given a set of {\color{red} [Speedup approach TBD]}

% \section{Performance Assessment}
\section{Experimental Setup}
\label{sec:performance}

%\input{Sections/groundTruth.tex}

%\subsection{Performance Assessment}

 \textbf{Ground Truth:} To the extent of our knowledge there is no relevant data available captured on a multi-domain environment. Thus, we follow the ground truth model and data generation process in~\cite{Vleeschauwer21_SLAdecomposition} to generate data for three domains.

 \textbf{Performance Assessment:} In order to assess the performance of the proposed approaches, we repeat the following process multiple times:

\noindent \textbf{\textit{Step 1.}} 
We generate $K$ partial SLAs per each of the $N$ domains $(\tau_{n,k}, \theta_{n,k})$ where the delay $\tau_{n,k}$ and the throughput $\theta_{n,k}$ are randomly sampled from a uniform distribution over pre-defined intervals. %$[\tau_{prop}, \tau_{max}]$ and $[0, \theta_{max}]$ respectively. 
We use the ground truth model~\cite{Vleeschauwer21_SLAdecomposition} for each partial SLA to determine the corresponding acceptance probability. Given this probability, we employ a coin toss to decide whether the partial SLA is accepted or not by the domain.
Sets of partial SLAs and the corresponding binary decision is used for learning.

%We randomly generate $K$ partial SLAs $(\tau_{n,k}, \theta_{n,k})$ per each of the $N$ domains and use the ground truth model to determine the probability that these requests are accepted. This probability is then used with a coin flipping to determine whether or not domain $n$ accepts it$(\tau_{n,k}, \theta_{n,k})$. %%This provides the data $(\tau_{n,k}, \theta_{n,k})$ and its corresponding feedback for learning risk models.
%The pairs of SLA $(\tau_{n,k}, \theta_{n,k})$ and its corresponding binary feedback are the data for learning.

\noindent \textbf{\textit{Step 2.}} We split the data in a training and a validation set, and train the  NN for each domain. Training ends when the loss on the validation set no longer decreases for $100$ epochs in a row.

\noindent \textbf{\textit{Step 3.}} With these trained risk models, we decompose a given E2E SLA using exhaustive search as in \cite{Vleeschauwer21_SLAdecomposition}, followed by SQP.

% \noindent \textbf{Step 4.} Finally, we present the decomposed SLA to the ground truth model and determine the probability (according to the ground truth model) that this decomposition $(\tau_n, \theta_n)$ is accepted in all domains, and hence, is globally accepted.
\noindent \textbf{\textit{Step 4.}} Finally, we feed the partial SLAs to the ground truth model to determine its acceptance probability per domain and estimate the E2E acceptance probability as the product of their individual values.

We average these E2E acceptance probabilities over all independent runs and compare it to the respective probability of the optimal decomposition, i.e., based on the ground truth. The E2E SLA we employ is $(\tau_{e2e}, \theta_{e2e}) = (100\mbox{ms}, 0.5\mbox{Gbps})$.

%\vspace{-0.5em}
\begin{figure}[t]
 \centering        
\captionsetup{justification=justified,singlelinecheck=off}
% \hfill
\subfloat[Ground truth]{\includegraphics[width=4cm]{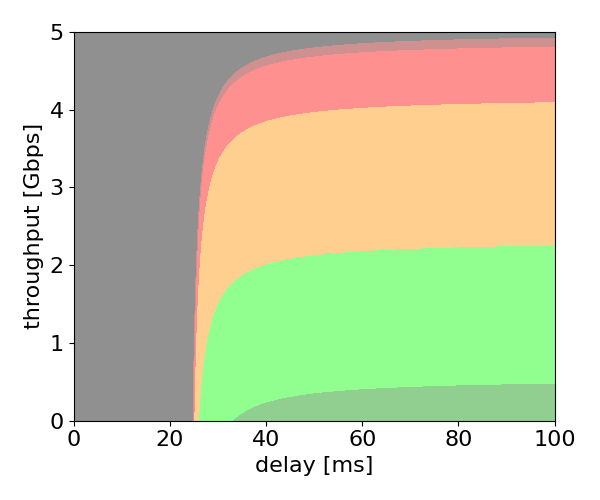}}
\hfill
e.g.\subfloat[Sample points]{\includegraphics[width=4cm]{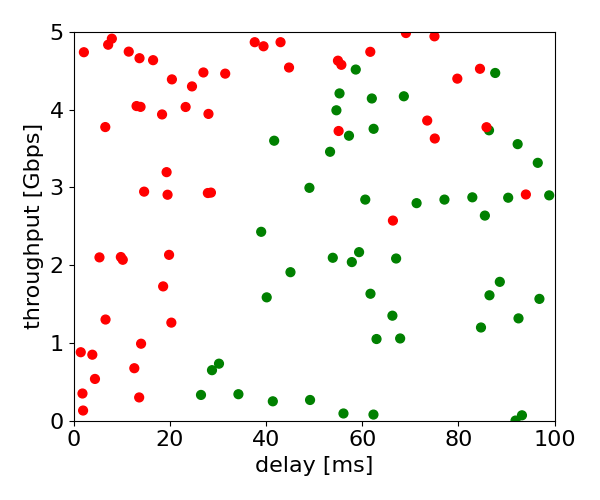}}
\hfill
\subfloat[Vanilla]{\includegraphics[width=4cm]{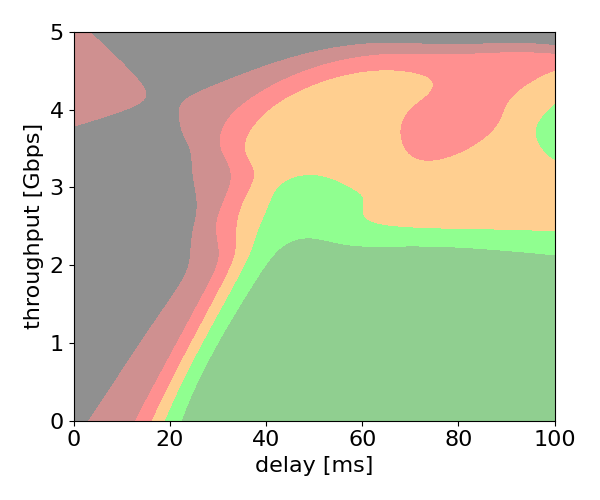}}
\hfill
\subfloat[Regularised]{\includegraphics[width=4cm]{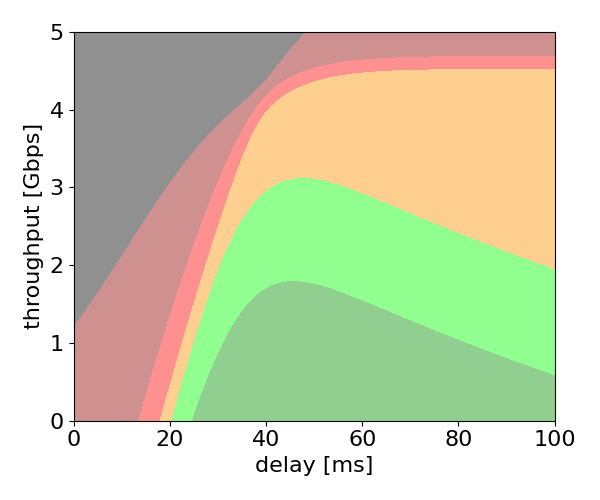}}
\hfill
\subfloat[AWET]{\includegraphics[width=4cm]{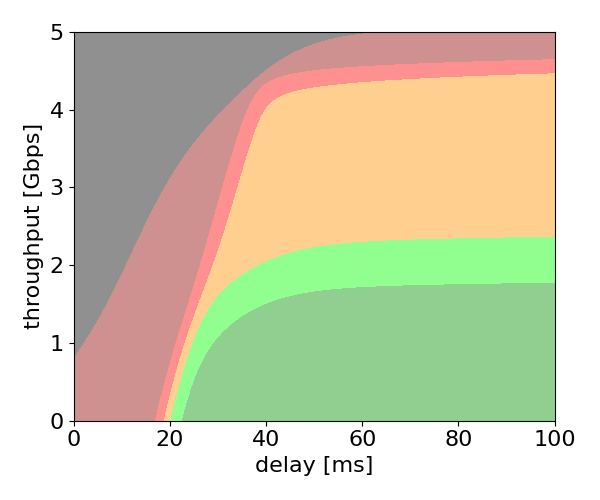}}
\hfill
\subfloat[CSE]{\includegraphics[width=4cm]{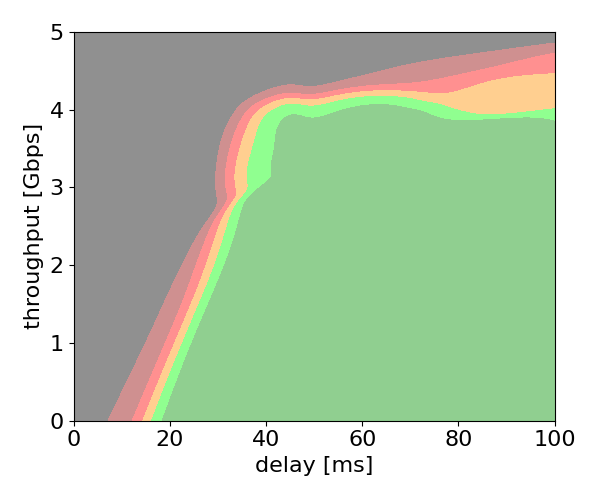}}
\hfill
\subfloat[PO]{\includegraphics[width=4cm]{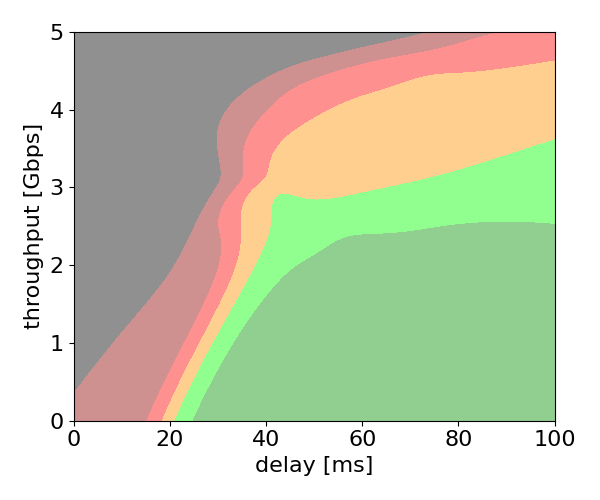}}
\hfill
\subfloat[MOL]{\includegraphics[width=4cm]{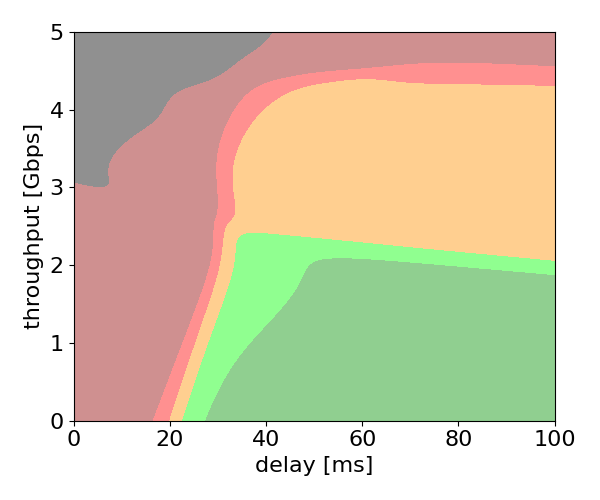}}
\hfill
\caption{Ground truth and learned risk models} \label{fig:contour}
\end{figure}

\begin{figure*}[ht]     
 \centering
\captionsetup{justification=justified,singlelinecheck=off}
    \subfloat[Avg. E2E acceptance probability.]{\includegraphics[width=.34\linewidth]{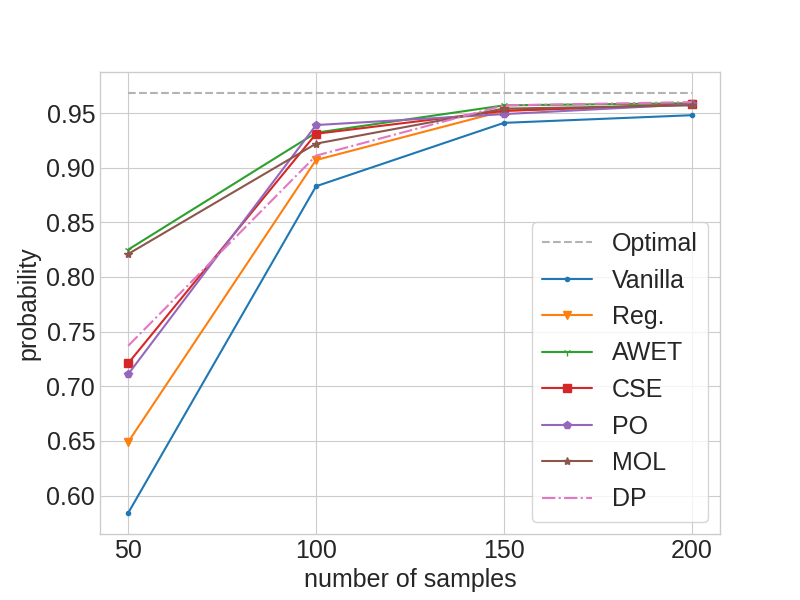}\label{fig:prob}}
    \subfloat[SD of E2E acceptance probability.]{\includegraphics[width=.34\linewidth]{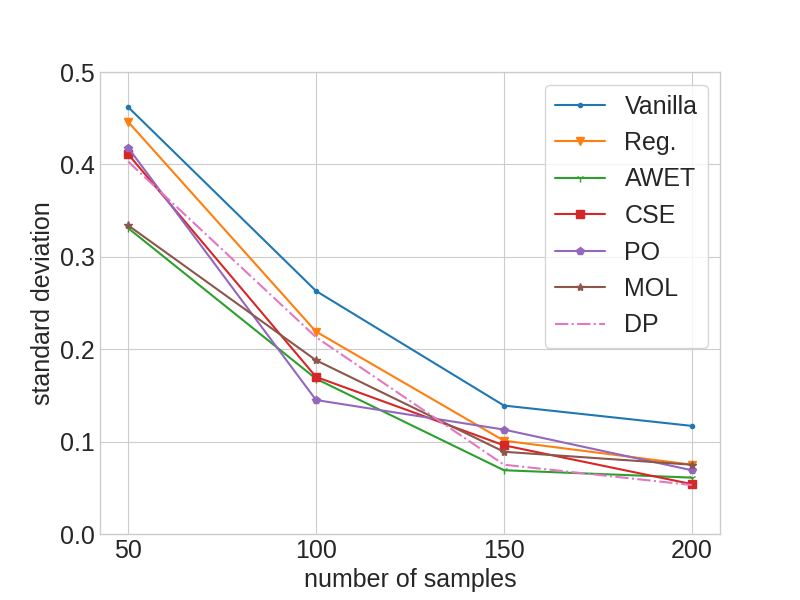}\label{fig:std}}
    \subfloat[Training time.]{\includegraphics[width=.34\linewidth]{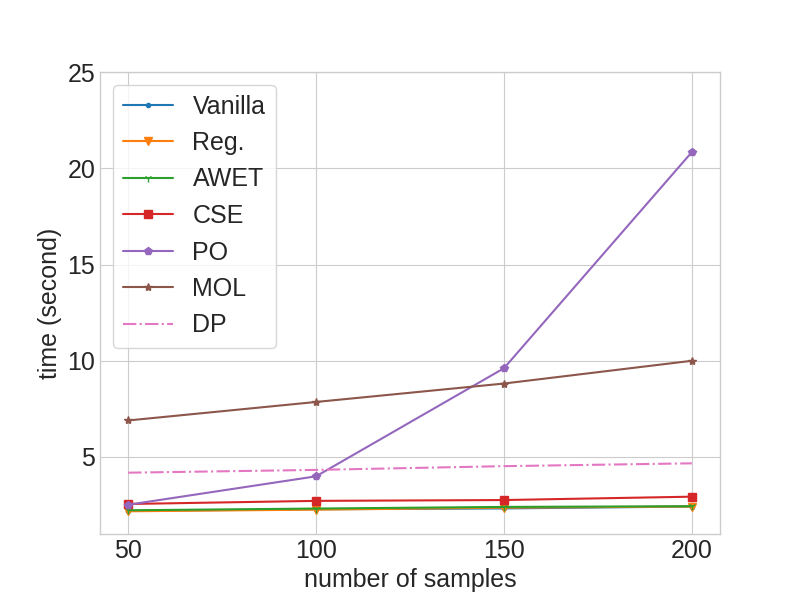}\label{fig:time}}
    % \justifying
    \caption{Performance of SLA decomposition.}\label{fig:perf}
\end{figure*}

%\subsection{Configurations}
%\textbf{Configurations:}
Hereafter, we describe the configurations for training hyper-parameters and NN architectures. To have a fair comparison, methods introduced in Section~\ref{sec:methods} are adapted on top of the same basic configuration. 
All experiments are conducted on a server with Intel Core i7-10700K CPU, 32 GB of RAM.
%, but with minor variations.% The setting of vanilla neural network is used as the basic configuration, and we merely show the difference for each method with respect to the basic configuration.
% \textbf{Vanilla Neural Network} The model is a simple 3-layer multi-layer perceptron (MLP), with 8 neurons each. The hyperbolic tangent (Tanh) activation function and batch normalization (BN)~\cite{https://doi.org/10.48550/arxiv.1502.03167} are applied for hidden layers in the order of linear-Tanh-BN. Remark that we disable the affine transformation in BN to avoid affecting the possible monotonicity of the neural network. The output layer uses sigmoid activation without appended BN. For the raining hyper-parameter, learning rate is set to $0.001$ and batch size is set to $32$. The objective function to minimize is in Equation~\ref{eq:bce}. After the risk models are obtained, we employ Sequential Quadratic Programming to solve the decomposition problem defined in Equation~\ref{eq:obj} and~\ref{eq:composition}, as suggested in~\cite{Vleeschauwer21_SLAdecomposition}.

% \textbf{Regularised Neural Network}
% We minimize the objective function in Equation~\ref{eq:reg}, where the balance factor $k$ is set to $0.1$.

% \textbf{Absolute Value Weight Transformation}
% The forward computation $w*x+b$ is replaced with $abs(w)*x+b$.

% \textbf{Mini-batch Order Loss}
% The objective to minimize is in Equation~\ref{eq:mol}, where the balance factor $k$ is set to $1.0$.

% \textbf{Conflicting Sample Elimination}

% \textbf{Probability Optimization}

%\vspace{-0.5em}
\noindent \textbf{\textit{Network Architecture}.} The base model is a 3-layer multi-layer perceptron (MLP), with 8 neurons each. The hyperbolic tangent (Tanh) activation function and batch normalization (BN)~\cite{ioffe2015batch} are applied for hidden layers in the order of linear-Tanh-BN. Remark that we disable the affine transformation in BN to avoid affecting the possible monotonicity of the NN. The output layer uses sigmoid activation without appended BN. The base model is adopted by all six methods.% For AVWT, an absolute value function is applied to weights before further multiplications, as mentioned in Section~\ref{sec:methods}.

% \begin{figure}[ht]
%     \centering
%     \includegraphics[width=1\columnwidth,scale=0.5]{Figures/arch.png}
%     \caption{Network architecture.}
%   \label{fig:arch}
% \end{figure}

\noindent \textbf{\textit{Hyper-parameters.}}
For training, the learning rate is set to $0.01$ and batch size to $16$. The dataset is split in $80\%$ for training and $20\%$ for validation. The objective function to minimize for each method is described in Section~\ref{sec:methods}. The balance factor for regularisation loss in Eq.~\ref{eq:reg} is set to $k=0.1$, and for the mini-batch order loss in Eq.~\ref{eq:mol} to $k=1.0$.  Both $k$ are determined by grid search in $\{0.01, 0.1, 1.0, 10.0\}$. To reduce statistical variability, all experiments are repeated $1000$ times, and average values are reported.% Note that for the first epoch of training, we temporary replace all labels in the dataset with $probability=0.5$ to eliminate the dependency on the weight initialization.

\section{Results}
\label{sec:results}

\subsection{Risk Model Evaluation}
%\nonindent \textbf{Risk Model Evaluation.}
%\textbf{Data Generation:}
The data (i.e. $x_i=(\tau_i, \theta_i)$, $y_i=1$ for acceptance, $y_i=0$ for rejection) are generated by the given ground truth model as described in Section~\ref{sec:performance} (Step 1). %as following:
%\begin{enumerate*}
 % \item For each data point $i$, the delay $\tau_i$ and the throughput $\theta_i$ are randomly sampled from the uniform distribution over the interval $[\tau_{prop}, \tau_{max}]$ and $[0, \theta_{max}]$ respectively.
 % \item The pair $(\tau_i, \theta_i)$ is fed to the given ground truth model to obtain the corresponding acceptance probability $p_i$.
 % \item Given $p_i$, a coin toss is employed to decide whether the SLA $(\tau_i, \theta_i)$ is accepted or rejected.% Finally, we assign $a_k=1$ in case of acceptance, otherwise $a_k=0$. Note that $a_k$ is a binary value.
%\end{enumerate*}
%\textbf{Contour Plot}
% Based on the ground truth model, we generate data for $3$ domains using the parameters defined in Table~\ref{tab:modelParameters}
%Data for $3$ domains are generated based on the aforementioned process.
%Various sample sizes (i.e., 50, 100, 150, 200) are tested in our experiments.
The contour plots for all learned risk models with arbitrary $100$ samples in a single domain are shown in Fig.~\ref{fig:contour}\{c-h\}, where the dark green, light green, orange, light red and dark red region correspond to an acceptance probability greater than $0.99$, $0.9$, $0.5$, $0.1$ and $0.01$, respectively. 
The gray region represents an acceptance probability lower than $0.01$. 
We examine the quality of the risk models by comparing these  plots to the ground truth model (Fig.~\ref{fig:contour}a).
 Fig.~\ref{fig:contour}b depicts the sampled points of our training data, where the green dots represent accepted requests while red dots rejected ones. From Fig.~\ref{fig:contour}\{d-h\}, we can observe that the proposed methods with the monotonicity constraint generate smoother decision boundaries than that of the vanilla NN in Fig.~\ref{fig:contour}c.
Vanilla NN in Fig.~\ref{fig:contour}c overfits the data with sampling error (the area with overlapping red and green dots in Fig.~\ref{fig:contour}b), which leads to irregular boundaries. AWET in Fig.~\ref{fig:contour}e obtains the most similar shape to the ground truth model, which is attributable to its strict monotonic constraint. CSE in Fig.~\ref{fig:contour}f shows the steepest gradient at the boundaries (i.e., moving from 0.99 to less than 0.01 acceptance probability), as the overlapping red and green dots in the sampled points are removed before training.

%\nonindent \textbf{SLA Decomposition.}
\subsection{SLA Decomposition}
% \textbf{Performance}
%To evaluate the performance of the decomposition based on the obtained risk models, we decompose an end-to-end SLA of $(\tau_{e2e}, \theta_{e2e}) = (100ms, 0.5Gbps)$ for 3 domains defined in~\cite{Vleeschauwer21_SLAdecomposition}, and then the decomposed SLA is presented to the ground truth model to calculate the acceptance probabilities regarding each domain. Finally the global acceptance probability is yielded by calculating the product of acceptance probabilities across 3 domains, given the assumption that all domains are statistically independent.
To evaluate the performance of the obtained  decomposition based on the risk models, we decompose an E2E SLA of $(\tau_{e2e}, \theta_{e2e}) = (100\mbox{ms}, 0.5\mbox{Gbps})$ using  the process described in Section~\ref{sec:performance} considering various sample sizes (i.e., 50, 100, 150, 200).
 We further include the method introduced in~\cite{liu2020certified} which imposes monotonicity by penalizing derivatives (DP) with respect to inputs.
Fig.~\ref{fig:prob} shows the results for the E2E acceptance probability. The dashed line indicates the theoretical optimum. The vanilla NN performs poorly especially for small sample sizes, while the rest methods are improved across all sample sizes. The AWET method outperforms all models, particularly for small sample sizes, as it guarantees monotonicity.
MOL method achieves similar performance to AWET, yet with notably longer training time (see Fig.~\ref{fig:time}). 
%However, as the sample size increases, methods with preprocessing (i.e., CSE and PO) outperform others due to the higher model expressiveness.
However, as the sample size increases, the discrepancy in performance becomes marginal for all methods.
Note that the balance factor $k$ in both regularised NN and MOL are not optimal. We believe their performance can be improved with further parameter search.
Fig.~\ref{fig:std} shows that the standard deviation (SD) of E2E acceptance probabilities for all methods decreases inversely to the sample size. Fig.~\ref{fig:std} in conjunction to Fig.~\ref{fig:prob} also indicates that when a model acquires larger average E2E acceptance probability has in general lower SD, which suggests higher robustness.

In Fig.~\ref{fig:time} we can see that the training time of PO exhibits quadratic growth, due to its extra constraint handling step, as stated in~\cite{Vleeschauwer21_SLAdecomposition}. The rest of the methods grow linearly over the number of samples, with the MOL having additional constant time and a larger slope for calculating pair-wise loss among samples, which is proportional to the square of batch size. 
 DP also requires additional constant time for introducing random samples during iterations of training~\cite{liu2020certified}.

\section{conclusion}
\label{sec:conclusion}
In this study, we propose an approach for decomposing an end-to-end SLA associated to a network slice request to the involved domains, adopting neural network-based risk models. The approach is applicable to any two-level network slice management system. The orchestrator is unaware of the state of each domain. The risk models are built using historical data pertaining to admission control. We propose six neural network-based approaches that utilize the monotonicity prior, such that the SLA can be adequately decomposed even with a small dataset. An empirical study on a synthetic multi-domain dataset demonstrates the efficiency of our approach.

% \appendix
% \input{Sections/groundTruth.tex}

% \input{Sections/acknowledgment.tex}

%\section*{Acknowledgment}

%The preferred spelling of the word ``acknowledgment'' in America is without  an ``e'' after the ``g''. Avoid the stilted expression ``one of us (R. B.  G.) thanks $\ldots$''. Instead, try ``R. B. G. thanks$\ldots$''. Put sponsor acknowledgments in the unnumbered footnote on the first page.

\bibliographystyle{IEEEtran}
\bibliography{References/references.bib}

\end{document}